\documentclass[%
  reprint,                % two-column "reprint" layout
  amsmath,amssymb,        % math packages
  aps,                    % style (aps/ pra/prl etc.)
  floatfix,               % better float placement
  superscriptaddress      % author addresses style
]{revtex4-2}

\usepackage[english]{babel}

\usepackage{amsmath}
\usepackage{graphicx}
\usepackage{bm}

\usepackage{natbib}
\begin{document}
%\begin{frontmatter}

% A changer parce que c'est moche
\title{Gravitational aggregation regimes: critical dissipation threshold, optimal rigidity and fractal transition}

%\author[INRAE]{Yohann Trivino}
%\affiliation[INRAE]{organization={IATE, INRAE},
%            addressline={Institut Agro, University of Montpellier, 34000}, 
%            city={Montpellier},
%            country={France}}

\author{Yohann Trivino}
\affiliation{Institut Agro, University of Montpellier, 34000, Montpellier, France}

\begin{abstract}
I present a three‐dimensional Discrete Element Method study of self‑gravitation and contact mechanics in cold granular assemblies, with direct Newtonian attraction between every particle pair and a linear visco‑elastic contact law. Particles are modeled as non‐cohesive spheres whose normal force is parameterized to reproduce a prescribed restitution coefficient, and rotations are integrated via quaternions to avoid singularities. By normalizing the stiffness $k_n$ by $k^*=G\,m^2/R^3$ and time by the free‐fall time, I perform systematic campaigns over dissipation ($\gamma_n$), stiffness ($\widetilde k_n=k_n/k^*$). My results reveal three aggregation regimes: for low $\gamma_n$, particles remain largely dispersive; above a critical $\gamma_n\approx5\times10^2$, aggregation accelerates until plateaus in $T_{\rm agg}/t_{\rm ff}$; and for stiffness $\widetilde k_n\sim10^6$, the aggregation time reaches a clear minimum. The cluster count $C/N_{\rm tot}$ likewise exhibits a non‐monotonic dependence on $\widetilde k_n$, with optimal cohesion at intermediate rigidity and peripheral isolation at extreme stiffness. Mapping the fractal dimension $F$ across $(\gamma_n,\widetilde k_n)$ further demonstrates transitions from compact ($F\approx3$) to ramified ($F\lesssim2$) structures. These findings quantify how microscopic contact laws govern both the kinetics and microstructure of gravity‐driven aggregation, providing a predictive framework for planetesimal formation and for calibrating DEM models against laboratory and micro‐gravity experiments.
\end{abstract}

\maketitle

%\begin{keyword}
%Discrete Element Method \sep  Self-gravity \sep Granular aggregation \sep Planetesimal formation
%\end{keyword}

%\end{frontmatter}

\section{Introduction}

In the context of the formation of solid bodies in protoplanetary disks, it is now accepted that the transition from a diffuse dust cloud to a cohesive aggregate results from a subtle balance between gravitational attraction and contact mechanics \cite{dominik1997physics, blum2008growth, okuzumi2012rapid}. In the early stages of micron-sized grain aggregation, turbulent agitation dominates the relative dynamics of particles, while mutual gravity only comes into play in the advanced stages of planetesimal formation \cite{blum2008growth, okuzumi2012rapid}. The formation of planetesimals begins with the coagulation of micrometric dust grains into fluffy, fractal aggregates. Several studies \cite{blum2008growth, kataoka2013static} have shown that these aggregates grow via Ballistic Cluster–Cluster Aggregation (BCCA) up to sizes of approximately $\sim1$ mm. Okuzumi et al. \cite{okuzumi2012rapid} further demonstrated that in collisions between aggregates of similar size, the fractal dimension remains close to $2$, and the internal density decreases significantly. This fractal growth phase continues until the impact energy becomes sufficient to compact the aggregates, a process that plays a critical role in the subsequent formation of planetesimals.

Only once sufficiently massive clouds of aggregates have formed does collective gravitational fall become predominant \cite{jansson2014formation}. Conventional studies are often based on N-body codes \cite{vacondio2021grand} or SPH methods \cite{manenti2019sph, rajab2021combined, bu2025numerical} to model the global collapse of instabilities, or on Discrete Element Method (DEM) simulations to describe granular dynamics under a uniform gravity field \cite{capasso2022application}; on the other hand, little effort has been devoted to integrating Newtonian particle-to-particle gravity within a three-dimensional DEM scheme, with a physically parameterized contact law and full rotation tracking.

Recent years have seen progress in applying discrete-element and non-smooth contact dynamics methods to self-gravitating granular systems. In particular, several studies using soft-sphere DEM and non-smooth contact dynamics (CD) have investigated collisional dynamics, aggregate formation, and small-body regolith mechanics (e.g., Sánchez \& Scheeres \cite{sanchez2011simulating} and others). These works have explored polydispersity, realistic contact models and self-gravity for applications to asteroidal surfaces and small-body formation. This work complements these efforts by performing a systematic parametric study of the influence of normal contact dissipation and contact stiffness on gravitational aggregation, using a spring–dashpot soft-sphere DEM calibrated on laboratory impact experiments. I position this contribution as an exploration of how normal dissipation and stiffness map to aggregation regimes and fractal morphologies, rather than as a claim of methodological novelty of the underlying DEM formulation.

%Historically, astrophysical DEM models have favoured simplified approaches in which the gravity applied is constant or linear \cite{tancredi2011granular, sunday2020validating, schwartz2012implementation, radjai2011discrete}, tangential and normal friction being calibrated empirically without any direct link to a coefficient of restitution measurable in the laboratory \cite{borykov2019empirical, luding2008cohesive}. Even if several previous studies have calibrated their interaction laws by comparing DEM simulations and experimental data \cite{seizinger2012compression, schrapler2012physics}, the coefficient of restitution $e$ strongly depends on the impact velocity \cite{schrapler2012physics} and is not constant. Similarly, most work explores monodisperse or 2D cases, omitting the crucial influence of actual polydispersity - already identified in the literature as a key factor in sorting and final porosity \cite{voivret2009morphology, schulze2015porosity}.

%To fill these gaps, 
I propose a model in which each pair of particles interacts according to Newton's law and, as soon as their spheres overlap, undergoes a damped linear contact force :

$$
F_{n,ij} = 
\begin{cases}
k_n\,\delta_{ij} \;-\;\gamma_n\,v_{n,ij} & \text{if }\delta_{ij}>0,\\
0 & \text{else},
\end{cases}
$$

with $\delta_{ij}=R_i+R_j-r_{ij}$ and $v_{n,ij}$ the relative normal velocity. The damping coefficient $\gamma_n$ is chosen to reproduce a fixed coefficient of restitution $e$ which, unlike heuristic parameterizations, allows direct control of collision anelasticity from experimental measurements. Gravity, meanwhile, is regularized by a softening $\varepsilon\ll R_{\min}$, ensuring a stable numerical transition between long-range gravitational interaction and contact.

Finally, I complement this formalism with quaternionic orientation tracking, ensuring that each particle retains a unit norm of rotation and that its moments of inertia remain consistent when tangential friction is added.

By deploying this model in systematic parametric campaigns - varying $\gamma_n$, the normalized stiffness $\widetilde k_n=k_n/(Gm^2/R^3)$ - I simultaneously address the cluster's aggregation rate, final compactness, porosity and fractal structuring. This approach provides a direct link between laboratory-validated contact laws and planetesimal physics, while offering a comparable framework for micro-gravity experiments \cite{wurm2021understanding, blum1995laboratory}, paving the way for cross-validation between numerical simulation and experimental study.

Following this introduction, Section 2 presents the complete formalization of this physical model and the numerical scheme implemented to ensure its accurate resolution. Building on this foundation, Section 3 examines the critical role of dissipation: I quantify its influence on particle dynamics and demonstrate how it modulates energy loss during collisions. Leveraging these insights, Section 4 explores the different aggregation regimes, highlighting the transition between loose assemblies and compact structures depending on interaction strength. I then continue, in Section 5, with an in-depth study of the fractal microstructure emerging from these aggregations, characterizing its fractal exponent and discussing its physical implications. Finally, Section 6 concludes this work by synthesizing key findings, emphasizing the perspectives opened by the results, and proposing directions for future developments.

\section{Method}
\label{sec:model}

\subsection{Physical Model}

\subsubsection{Geometric and mechanical properties}

Consider a set of $N$ rigid particles, each modeled by a sphere of radius $R_i$, mass $m_i$ and density $\rho$. In order to simulate aggregates that may be very heterogeneous (e.g., a cloud of asteroids or beads of various sizes), I allow $m_i \;\neq\; \frac{4}{3}\,\pi\,\rho\,R_i^3,$
i.e. each particle has an arbitrarily fixed mass $m_i$ and radius $R_i$. I then define an effective density $\rho_i = \frac{3\,m_i}{4\,\pi\,R_i^3}$. This freedom makes it possible to modulate the gravitational behavior relative to the contact: a denser particle (for the same radius) will undergo more mutual gravitational force in proportion to $m_i$, while its inertia at the collision does not change (because $I_i \propto m_i\,R_i^2$).

Each particle possesses $\bigl(\boldsymbol{r}_i(t),\,\boldsymbol{v}_i(t),\,q_i(t),\,\boldsymbol{\omega}_i(t)\bigr),$ where $\boldsymbol{r}_i\in\mathbb{R}^3$ and $\boldsymbol{v}_i=\dot{\boldsymbol{r}}_i$ are position and linear velocity, $q_i\in\mathbb{S}^3$ is the unit quaternion describing orientation, and $\boldsymbol{\omega}_i$ is angular velocity. In the following, I will briefly go over these standard definitions to emphasize the choices specific to this model coupling mutual gravitation and linear elasto-amortization contact.

\subsubsection{The law of universal gravitation and its regularization}

%For two particles $i$ and $j$, the mutual gravitational force is written as :

%\begin{equation}
%\begin{aligned}
%    \boldsymbol{F}_{g,\,ij} &= G\,\frac{m_i\,m_j}{\bigl(r_{ij}^2 + \varepsilon^2\bigr)^{3/2}}\,(\boldsymbol{r}_j - \boldsymbol{r}_i),\\
%    r_{ij} &= \bigl\|\boldsymbol{r}_j - \boldsymbol{r}_i\bigr\|.
%\end{aligned}
%\end{equation}
%I systematically introduce a small softening parameter $\varepsilon\ll R_{\min}$ to avoid singularity when $r_{ij}\to 0$. In practice, I choose $\varepsilon = \alpha\,R_{\min}$ with $\alpha\approx10^{-3}$, where $R_{\min}=\min_k R_k$. This regularization ensures that the gravitational force remains finite even in the case of overlap $r_{ij}<R_i+R_j$, simplifying the harmonization between gravitational shuttles and contact forces.

For numerical stability the Newtonian pairwise gravitational force is regularized by a small softening length $\varepsilon$. I define $R_{\min} = \min_k R_k$ (the smallest particle radius in the system) and set $\varepsilon = \alpha R_{\min}$ with a default $\alpha = 10^{-3}$ (this value can be adjusted in the input). The gravitational magnitude between particles $i$ and $j$ is then computed as :
\begin{equation}
    \boldsymbol{F}_{g}^{ij} = G\,\frac{m_i\,m_j}{\bigl(r_{ij}^2 + \varepsilon^2\bigr)}
\end{equation}
which prevents numerical singularities when center-to-center distances become small. The softening is a numerical measure and does not replace contact mechanics; the normal contact repulsion is still computed whenever overlap occurs.

\subsection{Numerical Scheme}

\subsubsection{Overlap and validity zone of the linear model}

%When two particles $i$ and$j$ are close enough for $r_{ij} < R_i + R_j,$ I define the overlap $\delta_{ij} =(R_i + R_j) - r_{ij}$. In this model, I treat only the $\delta_{ij} \ll R_i, R_j$, so that the linear approximation of contact remains reasonable. If $\delta_{ij}$ becomes too large (for example $\delta_{ij} > 0.1\,\min(R_i,R_j)$), the configuration is considered to be in collision: a maximum threshold is then applied $\delta_{\max}$ such as :
%
%$$
%F_{n,\,ij} = k_n\,\delta_{\max} - \gamma_n\,v_{n,\,ij}
%\quad\text{if}\quad \delta_{ij} \ge \delta_{\max},
%$$
%to avoid excessive thrusts that no longer make physical sense (as the geometry of the particles diverges from the rigid sphere). I typically choose $\delta_{\max} = 0.05\,R_{\min}$. The threshold $\delta_{\max}=0{,}05\,R_{\min}$ is chosen to guarantee $\delta\ll R$ and limit non-physical overlaps \cite{luding2008cohesive}. I use a linear damped spring model ("Spring-Dashpot") for its numerical simplicity. This approach, which is standard in DEM \cite{cundall1979discrete, li2022numerical}, qualitatively captures energy dissipation in shocks. Although the Hertz-Mindlin law is more realistic for elastic spheres, the linear model allows for easy adjustment of parameters (stiffness, damping) to reproduce target restitution values. Comparative studies show that for small overlaps, the overall dynamics remain similar.

Contact detection is performed when $r_{ij} < R_i + R_j$. For small overlaps, I use a linear elastic-dashpot contact law; the normal force is
$$
F_{n} = k_n\delta - \gamma_nv_n
$$
where $\delta$ is the geometric overlap, $v_n$ the normal relative velocity, $k_n$ the normal stiffness and $\gamma_n$ the normal damping coefficient. To ensure the small-deformation assumption remains valid I clamp $\delta$ with $\delta_{\max} = 0.05R_{\min}$ (overlaps exceeding this threshold are set to $\delta_{\max}$ and the case is flagged as numerically extreme). The spring–dashpot parameters $k_n$ and $\gamma_n$ are calibrated together to obtain a desired contact duration $t_{\rm contact}$ and effective restitution $e$ for the velocity range of interest (this is a standard practical approach in DEM \cite{cundall1979discrete, di2005improved, luding2008cohesive})

\subsubsection{Stiffness $k_n$ and damping coefficient $\gamma_n$ settings}

Rather than arbitrarily fixing $k_n$ and $\gamma_n$, I relate these two parameters to the coefficient of restitution $e_{ij}$ and the natural period of the contact oscillator:

\begin{equation}
\begin{aligned}
    m^*_{ij} &\;=\; \frac{m_i\,m_j}{m_i + m_j}, \\
    \omega_{0,\,ij} &= \sqrt{\frac{k_n}{m^*_{ij}}},
\end{aligned}
\end{equation}

\begin{equation}
\begin{aligned}
    \gamma_{n,\,ij} &= 2\,m^*_{ij}\,\eta_{ij}, \\ 
    e_{ij} &= \exp\Bigl(-\,\eta_{ij}\,\pi/\sqrt{1 - \eta_{ij}^2}\Bigr),
\end{aligned}
\label{eqn:rest}
\end{equation}
where $\eta_{ij}\in(0,1)$ is the pair's dimensionless damping factor $(i,j)$. In practice, I choose a unique value $\eta$ for the whole system, then I deduce :

\begin{equation}
\begin{aligned}
    k_n &= m^*_{\min}\,\omega_{0}^2 \\
    \gamma_n &= 2\,m^*_{\min}\,\eta,
\end{aligned}
\end{equation}
with $m^*_{\min} = \min_{i<j} m^*_{ij}$ and $\omega_{0} = \frac{\pi}{t_{\text{contact}}}\sqrt{1-\eta^2}$. The desired contact time $t_{\text{contact}}$ (typically $10^{-3}$ to $10^{-4}$s) fixes $\omega_{0}$. This choice ensures that the most brutal collision (the one involving the smallest reduced mass) is resolved with coherent damping. For two particles of greater reduced mass, their $\omega_{0,\,ij}$ will be lower, which slows the oscillation slightly, but the system remains stable.

\subsubsection{Detailed formulation of the normal force}

When $\delta_{ij}>0$, I define :

\begin{equation}
\begin{aligned}
    \boldsymbol{n}_{ij} &= \frac{\boldsymbol{r}_j - \boldsymbol{r}_i}{r_{ij}}\\
    v_{n,\,ij} &= (\boldsymbol{v}_j - \boldsymbol{v}_i)\cdot \boldsymbol{n}_{ij}.
\end{aligned}
\end{equation}
The modulus of the contact force can then be written as:

\begin{equation}
    F_{n,\,ij} = \underbrace{k_n\,\delta_{ij}}_{\text{elastic}} \;-\; \underbrace{\gamma_n\,v_{n,\,ij}}_{\text{dissipative}}.
\end{equation}
Assuming contact without tangential friction, the vector force applied to particle $i$ is

\begin{equation}
\begin{aligned}
    \boldsymbol{F}_{\text{contact},\,i\leftarrow j} &= -\,F_{n,\,ij}\,\boldsymbol{n}_{ij}\\
    \boldsymbol{F}_{\text{contact},\,j\leftarrow i} &= +\,F_{n,\,ij}\,\boldsymbol{n}_{ij}.
\end{aligned}
\end{equation}
When $F_{n,\,ij}<0$ (the case where damping prevails and rapid separation occurs), I impose the non-adhesion condition: I set $F_{n,\,ij} = 0$ if the algebraic expression leads to an attractive force. A tangential friction term with modulus $F_t = \mu\,F_n$ can play a significant role.

\subsubsection{Time-step selection}

I select the integration time-step $d_t$ to satisfy the stability/accuracy criterion
\begin{equation}
    d_t \leq t_{\rm contact}/20, \quad t_{\rm contact} \approx \pi \sqrt{\frac{m_{\rm eff}}{k_n}}
\end{equation}
where $m_{\rm eff}$ is a representative pair reduced mass (I use the smallest-particle-pair approximation $m_{\rm eff} \approx \min(m_i)/2$). Simulations violating this criterion are documented and a warning is emitted by the code.

\subsubsection{Friction}

In the present study the majority of simulations were performed with tangential friction coefficient $\mu = 0$ in order to isolate the influence of normal dissipative mechanisms. When $\mu > 0$ is used, tangential forces are computed using an incremental tangential displacement model with a Coulomb yield criterion ($F_t \leq \mu F_n$), and history of tangential displacement is maintained for each contacting pair.

\subsection{Model calibration (MEDEA experiments)}

%Several experiments have studied the aggregation of spherical particles in microgravity or free fall, providing data comparable to the output of a DEM model \cite{blum2001drop, blum1995laboratory, weidling2012free}. For example, \citet{blum2001drop} used a free-fall tower with fractal aggregates of monodisperse micron silica grains. They observed that, for the lowest impact velocities, aggregates stick together without restructuring, that compact structures form at intermediate velocities, and that high-speed collisions cause fragmentation \cite{blum2001drop}. This experiment uses monodisperse $(\sim\mu m)$ $SiO_2$ spheres in a gas-poor environment. It has enabled us to measure a fragmentation threshold (of the order of $\sim cm/s$) and qualitatively confirms the dependence of the outcome of collisions (adhesion, compaction, fragmentation) on the relative speed of the aggregates.

The comparison with the MEDEA drop/impact experiments is used as a calibration step rather than an absolute validation. Specifically, $\gamma_n$ (via the damping parameter $\eta$) and the contact-time setting were adjusted to reproduce the measured coefficient of restitution for the impact velocities tested in MEDEA. Only a small set of parameters (primarily $\gamma_n$) was tuned; all other parameters follow experimental grain properties. I quantify the sensitivity of the calibration in Appendix A (showing $\pm$ variations that still remain compatible with experimental scatter).

For a comparison based on contact physics, an interesting statistical measure is the ratio of the number of collisions leading to adhesion to all collision events. The MEDEA experiment (Weidling et al. \cite{weidling2012free}) consists of an agitation mechanism with an eccentric wheel and a glass vacuum chamber, which is connected to the vacuum system. In their work, Weidling et al. \cite{weidling2012free} track this ratio of adhesive collisions. I call a collision "adhesive" when the particles remain in contact after impact (equivalent to zero total restitution). In other words, these are impacts with a restitution coefficient $e=0$ (no rebound). The fraction of adhesive collisions is therefore the number of these impacts divided by the total number of simulated collisions. In this model without van der Waals forces, this fraction simply results from the high damping chosen for certain cases.

To reproduce the experiment, I use a consistent set of parameters with $0.5-1.5mm$ polydisperse silica grains (I choose here the mean size of $1$mm and a standard deviation of $0.25mm$) with a density $\rho = 2600 kg/m^3$ in a very high vacuum environment $(P\sim 3Pa)$ during a microgravity experiment (free-fall lap, duration $\sim4-5$ s) with an observed mean restitution $e \simeq 0.28$ at $v\sim8$ cm/s. The masses are calculated at constant density $\rho_p$ according to $m_i = \tfrac{4}{3}\pi R_i^3 \rho_p$. Thus, the mass distribution directly follows that of the radii.

\begin{table*}[tbh]
    \centering
    \small
    \begin{tabular}{|c|c|c|c|}
        \hline
        Parameter & Symbol & Value & Unit \\
        \hline\hline
        Grain radius & R & $10^{-3}$ & m\\
        Grain mass & m & $\frac{4}{3} \pi\rho R^3 \simeq 4.6 \times 10^{-6}$ & kg\\
        Gravitational constant & G & $6.674 \times 10^{-11}$ & $m^3.kg^{-1}.s^{-2}$\\
        Target restitution coefficient & e & 0.28 & - \\
        Damping factor adim. & $\eta$ & $\simeq 0.376$ (resolution via eq.\ref{eqn:rest}) & - \\
        Desired contact time & $t_{contact}$ & $10^{-4}$ & s \\
        Contact stiffness & $k_n$ & $m^*/\omega_0^2 = (m/2)\times(\pi/(t_{contact} \sqrt{1-\eta^2})))^2 \simeq 2.6 \times 10^3$ & N/m \\
        Damping & $\gamma_n$ & $2 m^* \eta = 2(m/2)\eta \simeq 1.7\times 10^{-6}$ & \\
        \hline
    \end{tabular}
    \caption{Set of parameters chosen to maintain the same sticking/rebonding behavior (same $e$) and contact physics comparable to the MEDEA experiment.}
    \label{tab:validation}
\end{table*}

\begin{figure}[!htb]
    \centering
    \includegraphics[width=0.95\columnwidth]{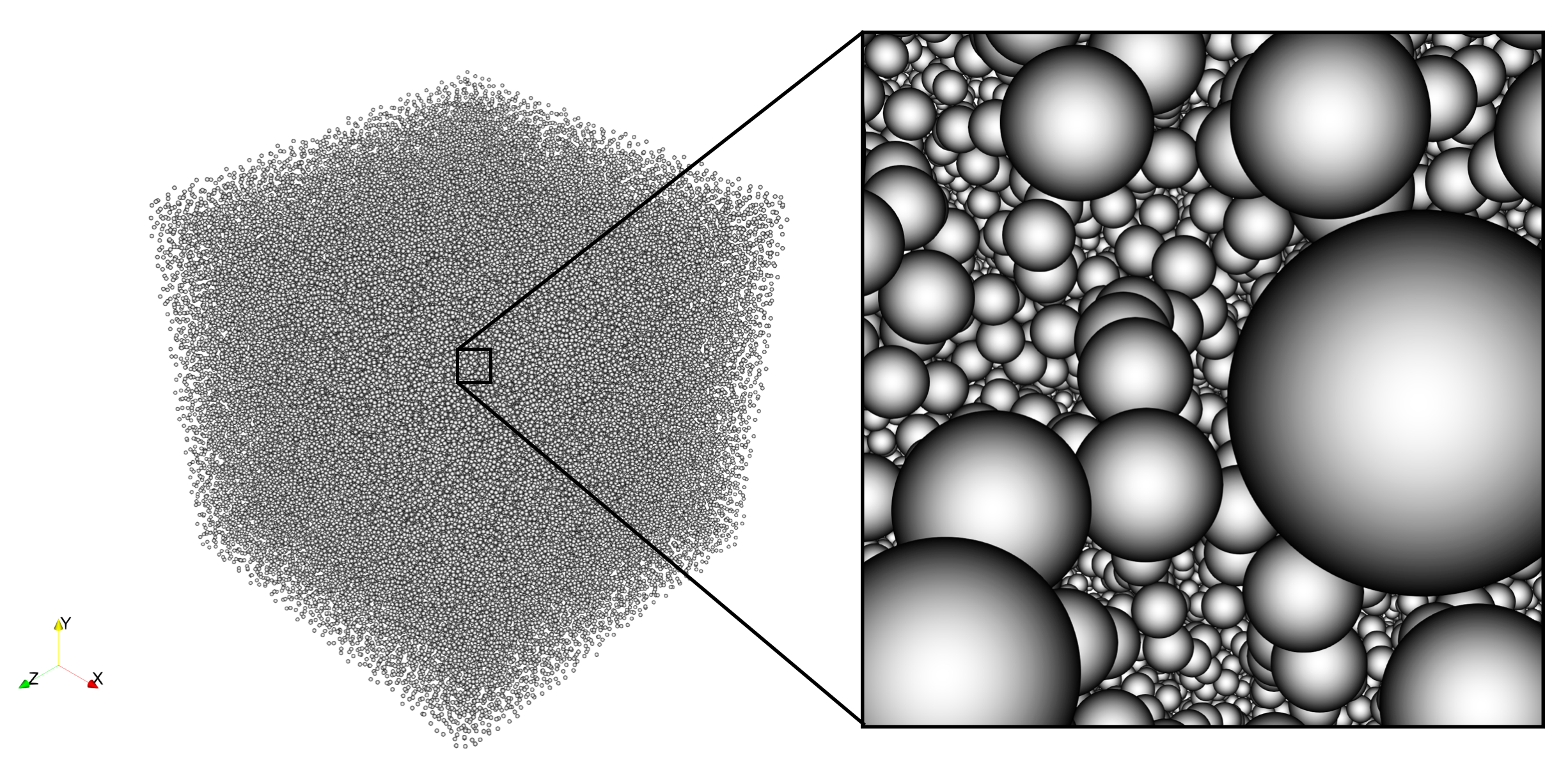}
    \caption{Initially loose sample of spherical particles designed to reproduce values produced by Weidling et al. \cite{weidling2012free}}
    \label{fig:validation_sample}
\end{figure}

With this set of parameters (Table \ref{tab:validation}), I can compare the contact physics of this model. The simulation (Fig \ref{fig:validation_sample}) recorded 129 adhesions out of 1867 collision events, equivalent to a fraction of colliding collisions of $\simeq 0.069$. The MEDEA experiment recorded 103 collisions, 7 of which resulted in adhesion, 95 in rebound and 1 in fragmentation. The experiment results in a fraction of colliding collisions $\simeq 0.068$. The small difference between the two results can be explained by the low number of collisions that the experiment can track.

\subsection{Simulation Protocol}

\begin{figure}[!htb]
\centering
\includegraphics[width=0.9\columnwidth]{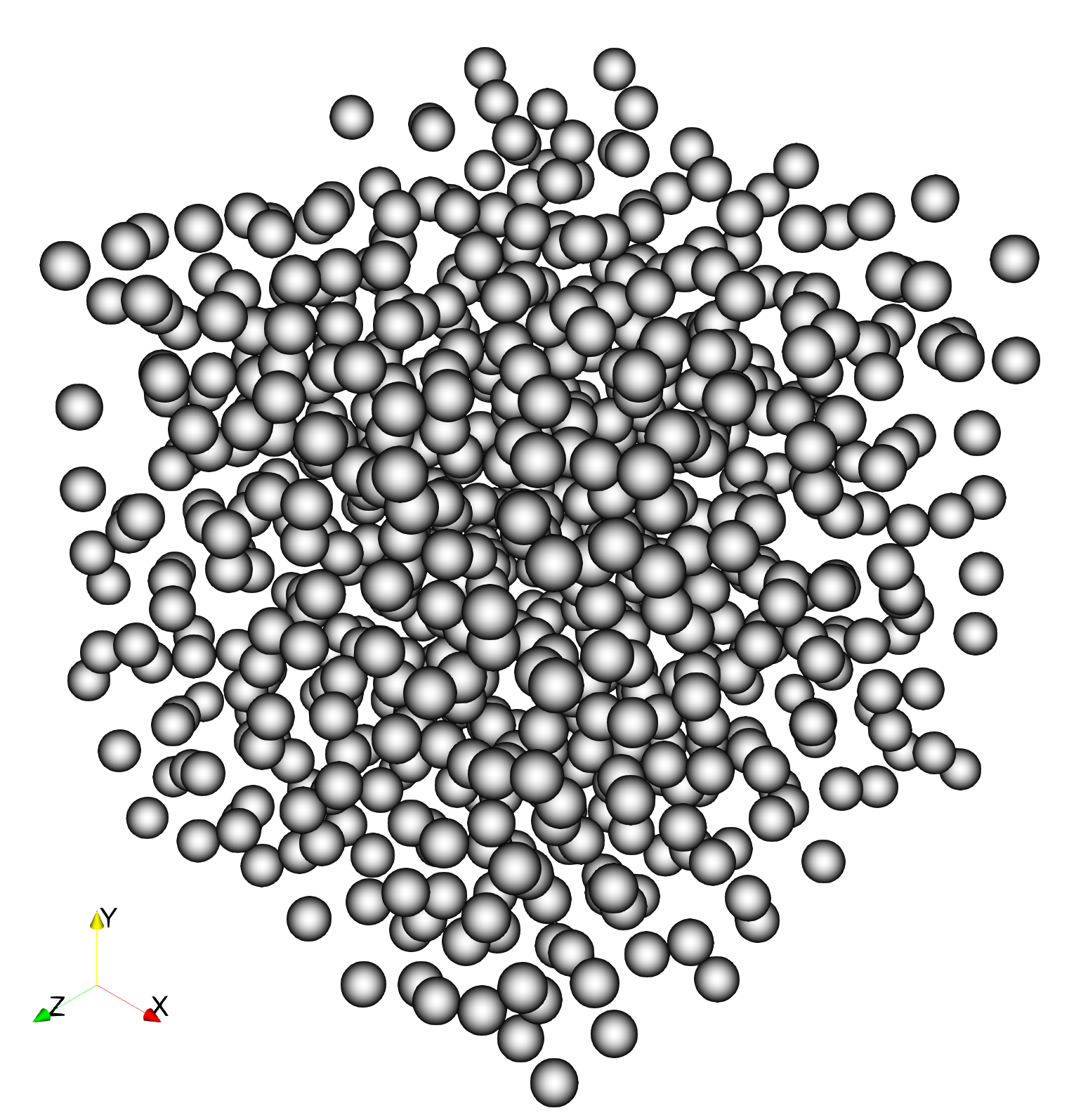}
\caption{An initially loose sample of 650 spherical particles distributed in a cubic domain of $\mathbb{R}^3$.}
\label{fig:box}
\end{figure}

Simulations involve preparing a loose, uniform sample of frictionless spherical particles inside a $\mathbb{R}^3$ cubic domain, then letting the laws of attraction guide the particles (Fig \ref{fig:box}). The simulations involve $650$ particles of radius $R$. Initially, the particles approach each other under the effect of attraction and may form an aggregate. I carried out $245$ simulations for values of the damping coefficient $\gamma_n$ between $[10,10^5]$ and the stiffness ratio $\widetilde k_n$ (defined eq.\ref{eqn:tildekn}) in $[10^5,10^8]$. I set the friction between particles to zero to eliminate frictional assumptions during analysis.

Using the numerical framework detailed above, I analyze in the following section the impact of dissipation mechanisms on the dynamic behavior of particles.

The values tested for normalized stiffness $\tilde k_n$ and damping $\gamma_n$ were chosen to cover a reasonable range of stiffnesses and restitutions. For example, $\gamma_n$ varies from 0 (no damping, $e=1$) to a value producing $e=0$ (completely inelastic impact). In the experiments in Blum et al. \cite{blum2008growth}, the coefficient of restitution for dust aggregates is typically in the range of $0.1-0.3$; the parameters therefore cover this range. Similarly, $\tilde k_n$ is set sufficiently large to limit recovery, while remaining commensurate with that used in other DEM studies \cite{ormel2009dust}. 

\section{The role of dissipation}
\label{sec:dissipation}

The aggregation time $T_{\rm aggr}$ is defined as the smallest instant (or simulation step) at which the mass fraction contained in the largest aggregate exceeds a threshold, typically 90$\%$ of the total particle mass. At each step, clusters are identified by neighborhood search: as soon as the size of the largest cluster $\ge0.9\,N_{\rm tot}$, the corresponding real time $T_{\rm aggr}$ is noted. So I will talk about aggregation only if I are in a regime where $T_{\rm aggr}$ is defined. 

The free-fall time $t_{\rm ff}$ gives the cloud's intrinsic gravitational time scale: first I calculate the initial center of mass, then I measure the maximum radius $R$ of the initial configuration (a particle's greatest distance from this center). If the total mass of the cloud is $M$, then

\begin{equation}
    t_{\rm ff} = \sqrt{\frac{R^{3}}{G\,M}}
\end{equation}
is the order of magnitude of the time required for gravitational collapse without internal pressure. The dimensionless quantity $\frac{T_{\rm aggr}}{t_{\rm ff}}$ therefore indicates whether aggregation occurs faster or slower than simple free fall. For the parameters in Table 1, $t_{ff}\approx 0.5$,s.

\begin{figure}[!htb]
\centering
\includegraphics[width=0.9\columnwidth]{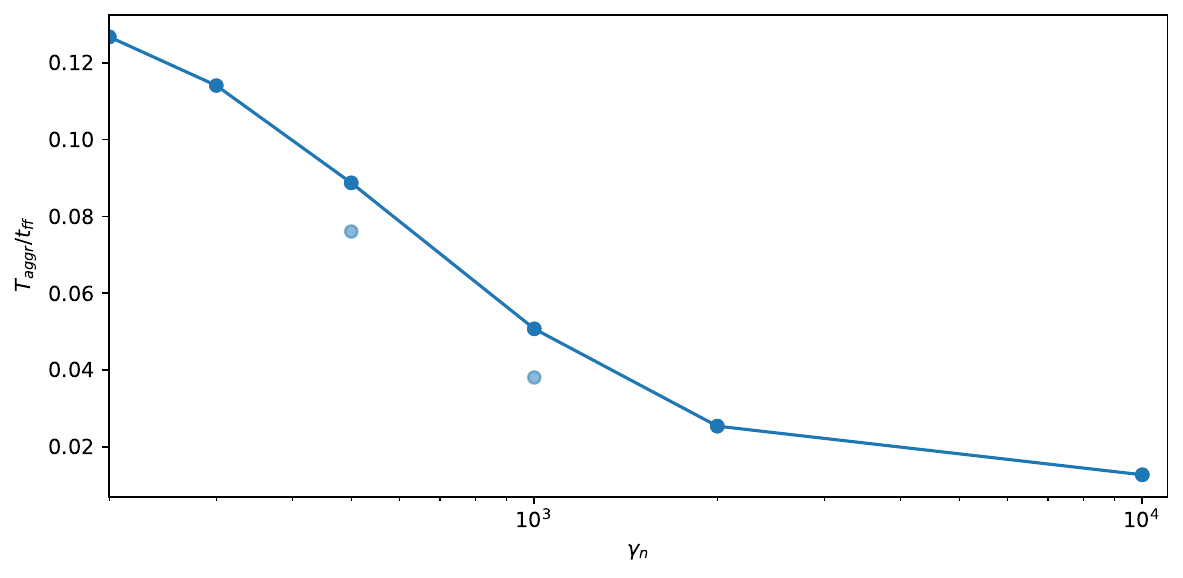}
\caption{Normalized aggregation time $\frac{T_{\rm aggr}}{t_{\rm ff}}$ as a function of damping coefficient $\gamma_n$ in simulations where aggregation is observed. Opaque circles: average achievements; transparent circles: individual raw samples.}
\label{fig:gamma_aggreg}
\end{figure}

Figure \ref{fig:gamma_aggreg} shows the normalized aggregation time $\frac{T_{\rm aggr}}{t_{\rm ff}}$ as a function of the damping coefficient $\gamma_n$ for tests where aggregation is observed. Aggregation time decreases with increasing $\gamma_n$.
This behavior can be explained: the higher the damping coefficient $\gamma_{n}$, the faster each impact between particles dissipates relative kinetic energy, and the less the particles bounce off each other. As a result, on first contact, a large proportion of their impact energy is dissipated (modelled by the $-\gamma_{n}\,v_{n}$ term in the contact law), gluing them together more effectively and accelerating the formation of a single aggregate.
In other words, for low values of $\gamma_{n}$, particles bounce for a long time before losing enough energy to remain bound, resulting in a long aggregation time $T_{\rm aggr}$. As $\gamma_{n}$ increases, dissipation is stronger, collisions become quasi-anelastic, and coalescence takes place in a reduced number of impacts: $T_{\rm aggr}/t_{\rm ff}$ falls.
Finally, beyond a certain threshold of $\gamma_{n}$, a plateau of reduced time can be observed: dissipation becomes so efficient that any further increase in $\gamma_{n}$ no longer significantly improves the speed of aggregation, as the collision are already practically sticky from the first impact. This monotonically decreasing evolution of $\frac{T_{\rm aggr}}{t_{\rm ff}}$ with $\gamma_{n}$ confirms that contact damping is the key parameter controlling the speed of aggregate formation under gravity.

But this curve alone does not tell us whether, by the time $T_{\rm aggr}$ appears to have been reached, there are still several sub-aggregates or whether all the particles have actually merged into a single cluster.
I define $C$ as the number of related components present at a given time in the simulation, and $N_{\rm tot}$ as the total number of initial particles. To determine $C$, I construct a neighborhood graph where two particles $i$ and $j$ are connected if their center-to-center distance $d_{ij}$ is less than the sum of their radii plus a small negligible tolerance. Dividing this number by $N_{\rm tot}$ gives the fraction $\frac{C}{N_{\rm tot}}$, which ranges from 1 (no aggregation, each particle isolated) to $\tfrac1{N_{\rm tot}}$ (all particles forming a single aggregate). %So I can talk about real or compact aggregation when $C/N_{\rm tot} < 0.1$. 

\begin{figure}[!htb]
\centering
\includegraphics[width=0.9\columnwidth]{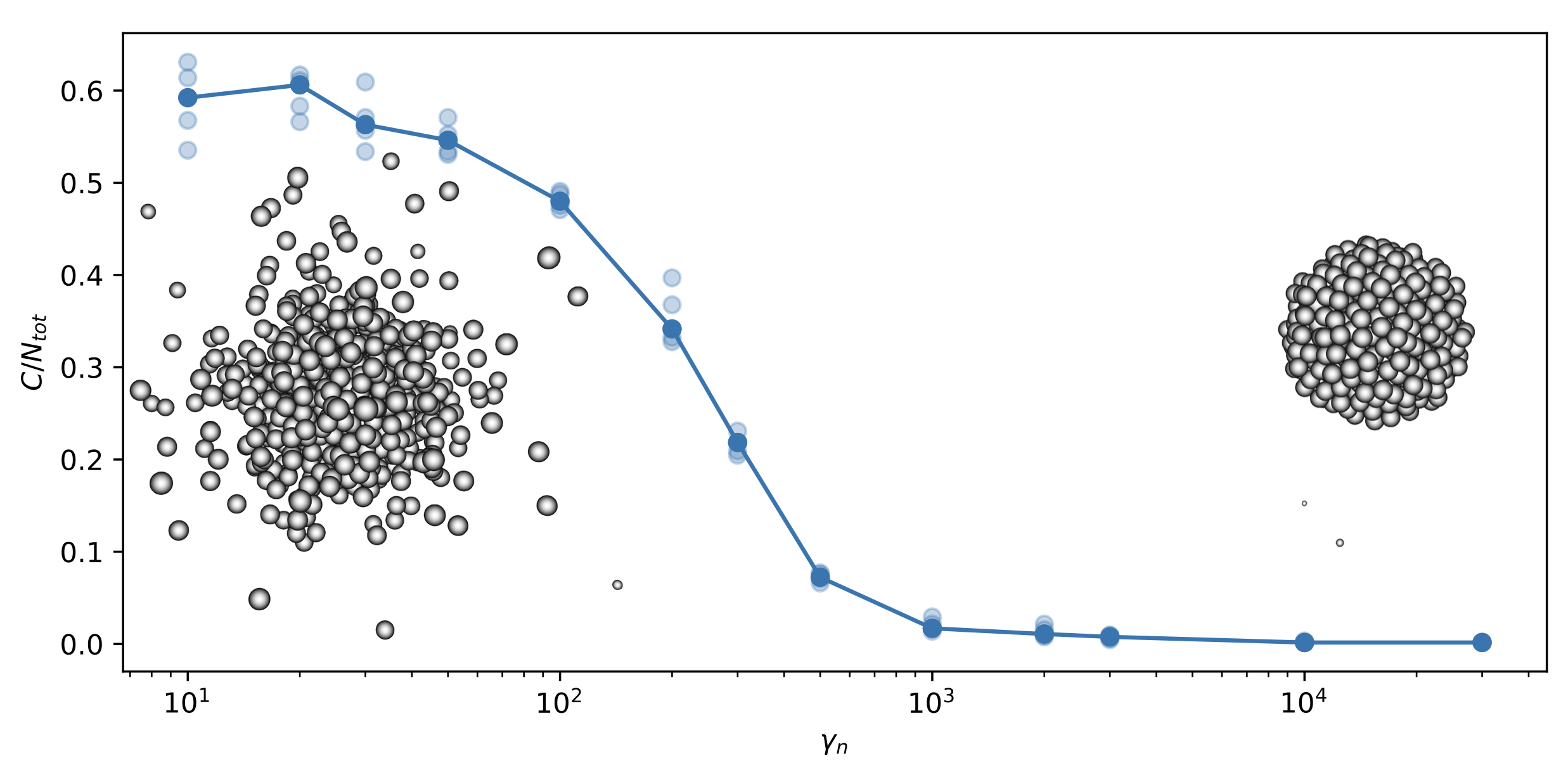}
\caption{Index $C/N_{tot}$ as a function of the damping coefficient $\gamma_n$. On the left, the final state of a simulation of the case $\gamma_n = 10$ and on the right a simulation with $\gamma_n = 10^4$. Opaque circles: average achievements; transparent circles: individual raw samples.}
\label{fig:gamma_cluster}
\end{figure}

Figure \ref{fig:gamma_cluster} shows the index $C/N_{tot}$ as a function of the damping coefficient $\gamma_n$, together with a visualization of two typical simulations. For low values of $\gamma_n$, the index $C/N_{tot}$ is $0.6$, implying that the majority of particles are isolated. As $\gamma_n$ increases, the index $C/N_{tot}$ logistically decreases to $0$ when large values of $\gamma_n$ are reached. As this figure shows, a zero $C/N_{tot}$ index implies a dense aggregate where all particles are bound together.  
This underlines the extent to which contact damping drives granular coalescence under gravity. When $\gamma_n$ is low, collisions are highly elastic and particles bounce back without losing enough energy to remain bound: I then observe an index $C/N_{\rm tot}\approx0.6$, i.e. most grains remain isolated or form only small aggregates. As $\gamma_n$ increases, each collision becomes increasingly anelastic and dissipates sufficient relative velocity to bring the particles together; the curve then adopts a logistic form, reflecting a rapid transition from a dispersed regime (many clusters) to a cohesive regime (few clusters). Beyond a certain critical threshold of $\gamma_n$, all kinetic energy is evacuated on first contact, and the index $C/N_{\rm tot}$ falls to zero: all particles end up in a single dense aggregate. This transition shows that there is a minimum value of dissipation beyond which the system switches abruptly from a granular gas to a compact cluster.

Despite aggregation, three possibilities are conceivable: the aggregate reaches a dynamic equilibrium, remains over-damped or remains too agitated. The evolution of the normalized index $C/N_{\rm tot}$ only provides information on the dispersion or coalescence of particles, without saying anything about how kinetic and potential energy balance within the cluster.
Let $\widetilde K(t)$ and $\widetilde P(t)$ be the dimensionless kinetic and potential energies, defined respectively by:

\begin{equation}
\begin{aligned}
    \widetilde K(t)&=\frac{E_{k}(t)}{\lvert E_{p}(0)\rvert}\\
    \widetilde P(t)&=\frac{E_{p}(t)}{\lvert E_{p}(0)\rvert}\,
\end{aligned}
\end{equation}
where the total kinetic energy and total gravitational potential at time $t$ are written :

\begin{equation}
\begin{aligned}
    E_{k}(t) &=\sum_{i=1}^{N}\tfrac12\,m_{i}\,\bigl\lVert \mathbf v_{i}(t)\bigr\rVert^{2}\\
    E_{p}(t) &=-\sum_{1\le i<j\le N}\frac{G\,m_{i}\,m_{j}}{\lVert \mathbf x_{i}(t)-\mathbf x_{j}(t)\rVert}
\end{aligned}    
\end{equation}

The ratio
\begin{equation}
    \frac{\widetilde K(t)}{\lvert \widetilde P(t)\rvert} = \frac{E_{k}(t)}{\lvert E_{p}(t)\rvert}
\end{equation}
is the index that provides real-time information on the system's energy balance.
I then introduce the normalized time $t_{norm} =\frac{t}{t_{\rm ff}}$ where $t$ denotes the actual time elapsed since the start of the simulation and $t_{\rm ff}$ is the gravitational free-fall time characteristic of the initial cloud. By dividing each observation time $t$ by this $t_{\rm ff}$, I obtain $t_{norm}$, which makes the time curves dimensionless and comparable from one simulation to the next.

\begin{figure}[!htb]
\centering
\includegraphics[width=0.9\columnwidth]{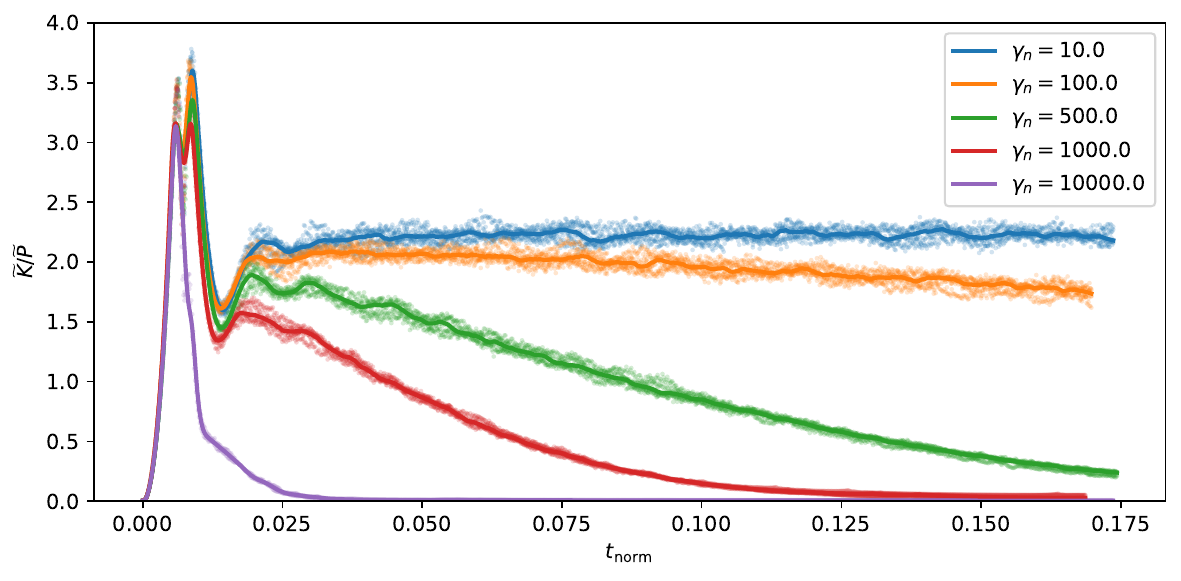}
\caption{Median evolution of the $\widetilde{K}/\widetilde{P}$ energy ratio over time for several $\gamma_n$ values. The transparent points represent the raw data for all simulations.}
\label{fig:gamma_t_virial}
\end{figure}

Figure \ref{fig:gamma_t_virial} shows the median evolution of the $\widetilde{K}/\widetilde{P}$ energy ratio over time for several $\gamma_n$ values. At the start of the simulations, for all values of $\gamma_n$, I observe a peak in $\widetilde{K}/\widetilde{P}$ due to the initial collision. Then, for all values of $\gamma_n$ except the largest $\gamma_n = 10^4$, a second peak is observed. After the first peak, the $\gamma_n = 10^4$ curve falls to $\widetilde{K}/\widetilde{P} = 0$. The other cases (with a lower $\gamma_n$) rise again and finally enter a more stable regime, either slowly decreasing to $0$ or remaining constant.
This evolution reflects the competition between the injection of kinetic energy during gravitational collapse and dissipation by collisions. At the very beginning, the "first collision" corresponds to the cold cloud's transition to rapid collapse: all simulations then show a peak of $\widetilde K/\widetilde P$ well above the virialization value (0.5), a sign of an abrupt exchange of kinetic potential.
For the maximum value $\gamma_n=10^4$, the normal damping is so high that, from this first collision, all the kinetic energy is immediately evacuated: $\widetilde K/\widetilde P$ falls back to almost zero and the system remains frozen, with no subsequent rebound or oscillation.

On the other hand, for $\gamma_n<10^4$, the damping is insufficient to immediately absorb the energy of the initial collision: I then observe the second peak, which corresponds to a collective rebound or internal reorganization of the particles (second arm of the dynamics) before dissipation takes over definitively. Finally, these intermediately-damped cases gradually enter a more stable regime: either their $\widetilde K/\widetilde P$ slowly decreases towards zero, or it remains suspended at a non-zero level, a sign that the system has reached a quasi-viralized state where residual kinetic energy remains around a constant fraction of potential energy.
In a purely theoretical framework, if damping $\gamma_n$ regulates the amount of energy dissipated at each contact, the contact stiffness $\widetilde k_n$ determines the duration and amplitude of these same collisions: a soft particle (low $\widetilde k_n$) dampens slowly and undergoes strong deformations, while a rigid particle (high $\widetilde k_n$) returns most of the kinetic energy in a brief impact. So, to predict not only at what velocity but also how grains bind or rebound, it is essential to couple the study of dissipation (via $\gamma_n$) with that of rigidity (via $\widetilde k_n$).
Having highlighted the role of dissipation, I now explore in section 4 the different aggregation regimes which emerge according to the parameters of rigidity and dissipated energy.

\section{Aggregation regimes}

In this section, the value of $\gamma_n$ is set to $5\times10^2$, which corresponds to the minimum value at which I can speak of a dense aggregate, according to the previous section.
The stiffness scale chosen is $k^*$, which reflects the order of magnitude of the gravitational forces at contact with a particle of mass $m$ and radius $R$. I start with the estimate

\begin{equation}
    F_{\rm grav}\sim\frac{G\,m^2}{R^2}
\end{equation}
for two identical grains at the point of contact. If I imagine a characteristic deformation $\delta\sim R$, the equivalent of a gravitational stiffness is written :
\begin{equation}
    k^* \;\equiv\;\frac{F_{\rm grav}}{\delta} \;\sim\;\frac{G\,m^2}{R^3}\,.
\end{equation}

I define the normalized stiffness:
\begin{equation}
    \widetilde k_n \;=\; \frac{k_n}{k^*} \;=\; \frac{k_n\,R^3}{G\,m^2}\,
\label{eqn:tildekn}
\end{equation}

This dimensionless parameter directly compares the mechanical contact stiffness $k_n$ with the stiffness scale imposed by gravity between two particles of the same size.

\begin{figure}[!htb]
\centering
\includegraphics[width=0.9\columnwidth]{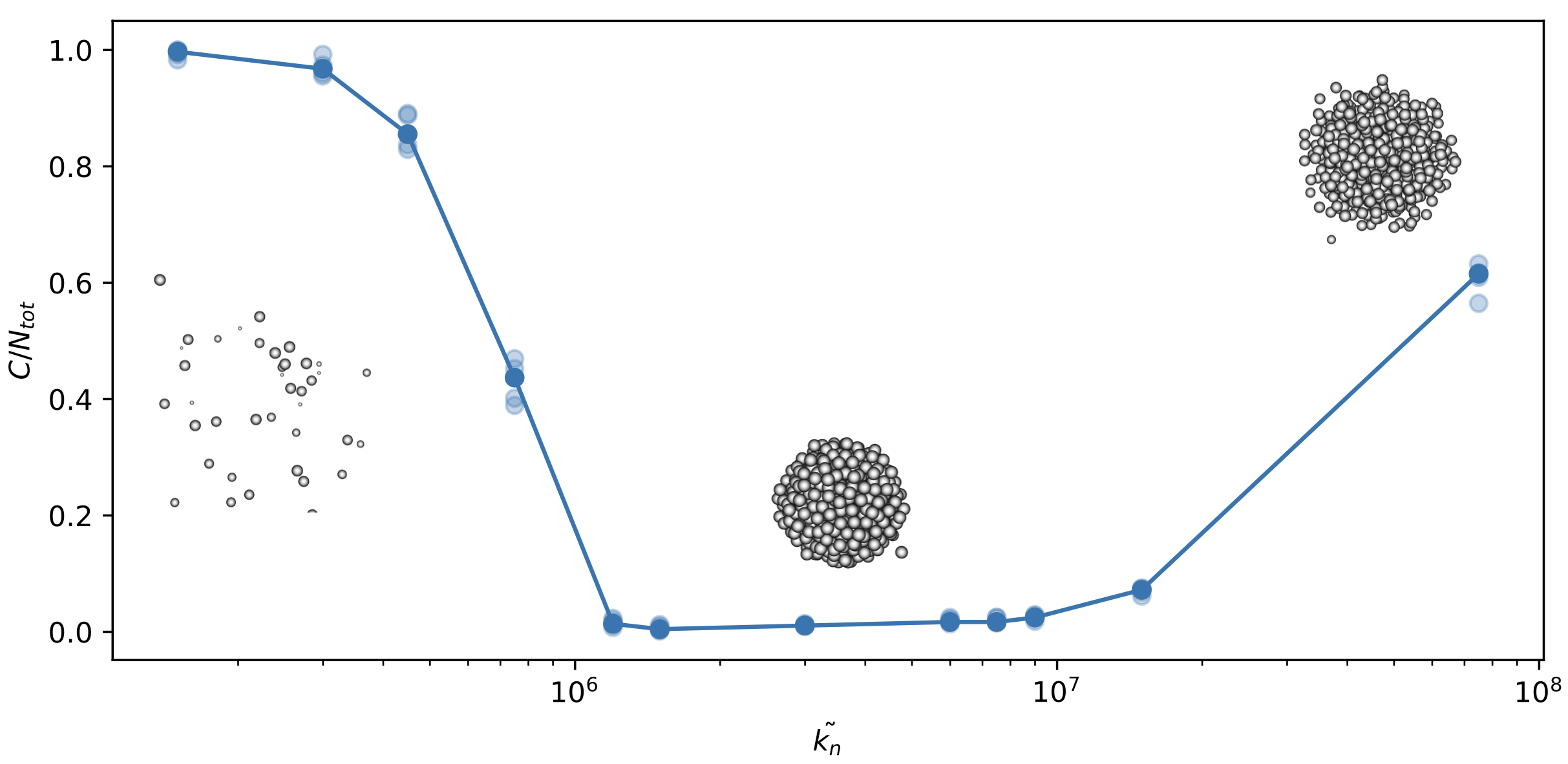}
\caption{Index $C/N_{\rm tot}$ as a function of $\widetilde k_n$. On the left, center and right, the final states of simulations proposing respectively $\widetilde k_n = 10^5$, $\widetilde k_n = 5\times10^6$ and $\widetilde k_n = 10^8$. Opaque circles: average achievements; transparent circles: individual raw samples.}
\label{fig:kn_cluster}
\end{figure}

Figure \ref{fig:kn_cluster} shows the index $C/N_{tot}$ as a function of $\widetilde k_n$, together with a visualization of typical simulations of the value of $\widetilde k_n$. For low values of $\widetilde k_n$, $C/N_{tot}$ is very large ($\simeq 1$), the particles are not aggregated and form a cloud of distant particles. The index decreases with increasing $\widetilde k_n$ until it reaches a value close to $0$. At these values of $\widetilde k_n$, the particles form a dense cluster. Then, as $\widetilde k_n$ increases further (from $\widetilde k_n = 10^7$), the $C/N_{tot}$ index increases again until it reaches around $0.6$. At this stage, the particles agglomerate, but some particles at the periphery remain disassembled.

The shape of this curve reveals three distinct physical regimes:

\begin{itemize}
    \item Soft regime ($\widetilde k_n < 10^6$): When the contact stiffness is extremely low compared to the gravitational scale ($\widetilde k_n \approx 1$ or less), particles sink strongly into each other on contact and experience significant overlap. The contact forces are then too weak to counteract the gravitational pressure, so the particles slide and disperse rather than bind, resulting in a $C/N_{\rm tot}\approx1$ index.\\
    \item Intermediate regime ($10^6 \lesssim \widetilde k_n \lesssim 10^7$): As stiffness increases, deformations decrease and contact forces become strong enough to resist and hold the grains together. This leads to a cohesive state: $C/N_{\rm tot}$ falls towards zero, indicating that a single dense aggregate is formed. This is the "sweet-spot" where contact rigidity is sufficient to consolidate the cloud without making the collisions excessively elastic.\\
    \item Regime too rigid ($\widetilde k_n \gtrsim 10^7$): Beyond a certain threshold, the contacts become practically indeformable and behave like highly elastic collisions: the particles bounce back instead of locking into place. In concrete terms, particles close to the surface of the main aggregate can no longer lock into place and remain isolated. The index $C/N_{\rm tot}$ then rises towards $\sim0.6$, reminding us that, despite a still agglomerated core, a significant fraction of peripheral grains remain unbound.
\end{itemize}

From an energetic point of view, the contact stiffness $\widetilde k_n$ controls both the local stability of the bonds (measured by the index $C/N_{\rm tot}$) and the rate at which kinetic energy is dissipated during collapse. In other words, a more rigid contact limits deformation and favors the rapid formation of a dense aggregate, but also risks prolonging coalescence processes by elastic rebound.

\begin{figure}[!htb]
\centering
\includegraphics[width=0.9\columnwidth]{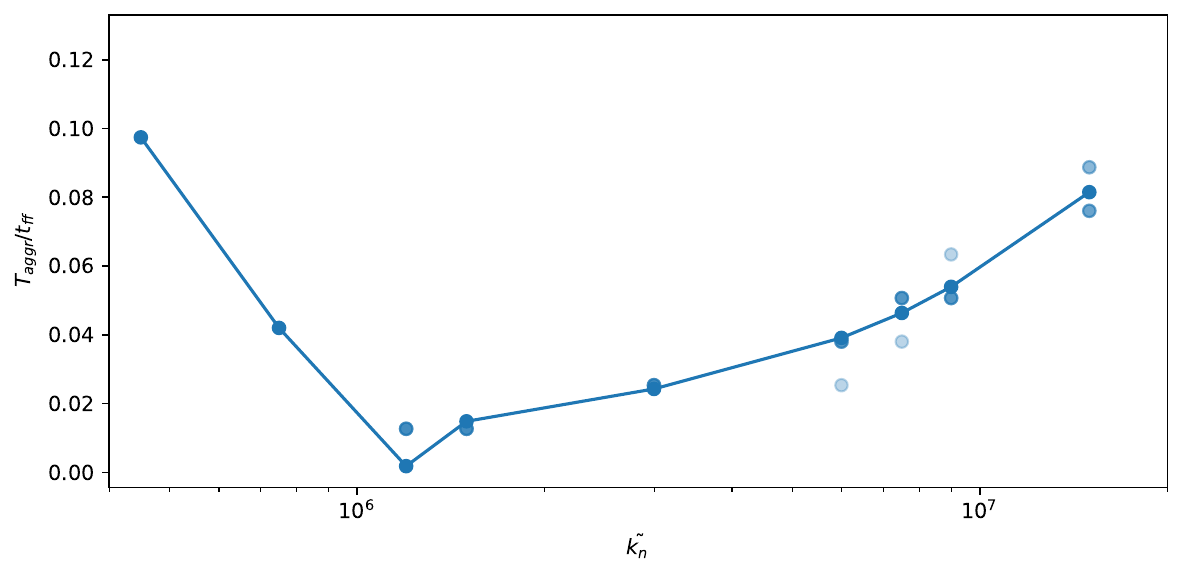}
\caption{Normalized aggregation time $\frac{T_{\rm aggr}}{t_{\rm ff}}$ as a function of stiffness $\widetilde k_n$, Opaque circles: average achievements; transparent circles: individual raw samples.}
\label{fig:kn_aggreg}
\end{figure}

Figure \ref{fig:kn_aggreg} shows the normalized aggregation time $\frac{T_{\rm aggr}}{t_{\rm ff}}$ as a function of stiffness $\widetilde k_n$. For low values of $\widetilde k_n$, the aggregation time decreases and then reaches a minimum at $\widetilde k_n = 10^6$. After this minimum, the aggregation time only increases with increasing $\widetilde k_n$. This highlights a stiffness optimum for the fastest formation of a gravitational aggregate. For $\widetilde k_n\ll10^6$, the contacts are too soft and the particles push excessively into each other without creating effective bonds. Collisions last a long time and produce only weak cohesion, so that $\tfrac{T_{\rm aggr}}{t_{\rm ff}}$ remains high. Around $\widetilde k_n\approx10^6$, contact rigidity reaches the ideal balance between minimum deformation and sufficient dissipation. In this regime, collisions dissipate just enough energy to rapidly bond the grains, while being soft enough to absorb the violence of the impact without exaggerated rebound. This is why the normalized aggregation time passes through a minimum: the cluster forms in a minimal number of steps, well below the pure free-fall time.

For $\widetilde k_n\gg10^6$, contacts become quasi-rigid and highly elastic: each collision is more like a rebound than a coalescence. The particles, although attracted, struggle to convert their kinetic energy into a lasting bond, which again lengthens $\tfrac{T_{\rm aggr}}{t_{\rm ff}}$. Successive collisions slow convergence towards a single aggregate, and the system takes longer to lose enough energy to bind.
This behavior highlights the existence of a mechanical optimum for the speed of aggregation, but what about the final microstructure of the cluster? In other words, does the value of $\widetilde k_n$ influence only the aggregation time, or does it also condition the nature of the internal arrangement? With too high a stiffness, agglomeration is slower and more incomplete, so can I expect looser, even fragmented structures? Only a more detailed geometric analysis (average coordination, tortuosity, local density, etc.) can answer this question.
The macroscopic characteristics of the aggregates being established, section 5 focuses on the detailed study of their fractal microstructure and its characteristic dimensions.

\section{Aggregate micro-structure}

The fractal character of aggregates has been extensively studied for aggregate growth in dusty environments \cite{blum2001drop, dominik1997physics}. Classical aggregation models produce open ramified structures (BCCA) or more compact clusters (BPCA) depending on the sticking dynamics and relative velocities. The fractal-dimension results place the aggregates of the intermediate/soft regimes in a regime compatible with ramified structures observed experimentally for adhesive dust and with theoretical predictions for ballistic cluster aggregation. I compare this numerical fractal exponents to experimental and theoretical values \cite{blum2001drop, dominik1997physics} and discuss the implications for aerodynamic settling and planetesimal accretion. I also stress finite-$N$ limitations: for significantly larger $N$ the cluster-cluster aggregation and long-range collective effects may change the morphology; addressing that requires $O(N log N)$ or $O(N)$ gravity solvers for future work.

\begin{figure}[!htb]
\centering
\includegraphics[width=0.9\columnwidth]{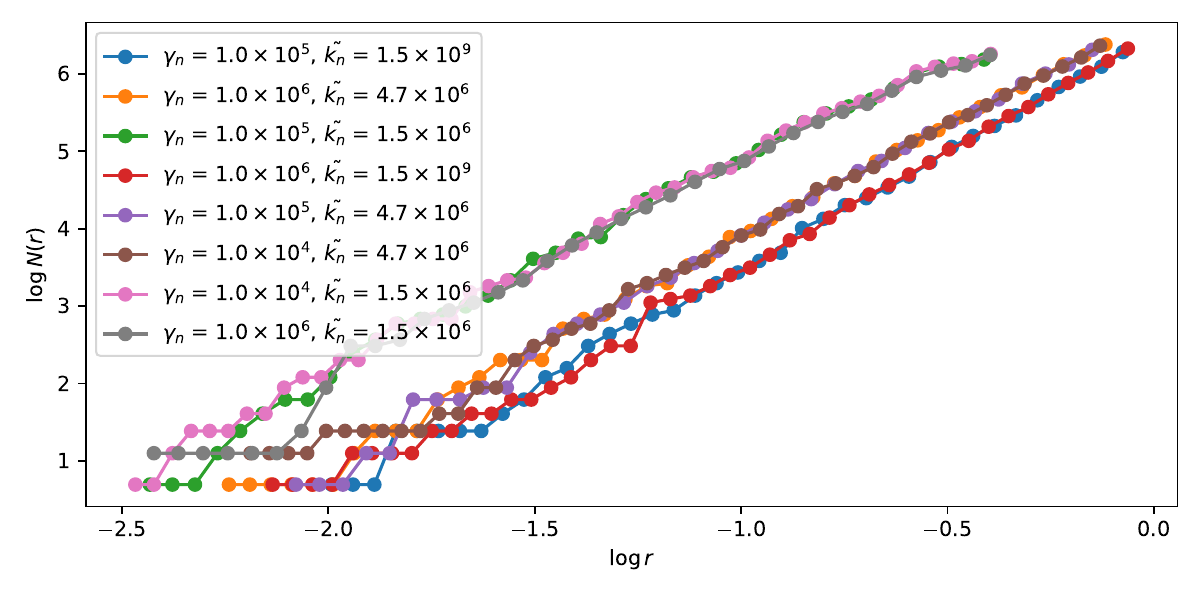}
\caption{Curves $\log N(r)$ vs $\log r$ for few simulations where $N(r)$ refers to the number of particles located at a distance less than or equal to $r$ from the center of mass. Each curve represents a different simulation.}
\label{fig:logN_minus_logr}
\end{figure}

I introduce the fractal dimension $F$ of the cluster, which I evaluate by plotting $\log N(r)$ as a function of $\log r$ and extracting the slope by linear regression. Here, $N(r)$ denotes the number of particles located at a distance less than or equal to $r$ from the center of mass. More precisely, if the cluster develops as an object of fractal dimension $F$, then I observe (in Fig \ref{fig:logN_minus_logr}) :
$$
N(r)\;\propto\;r^{F},
$$
and the slope of the straight line $\log N(r)$ vs. $\log r$ directly provides the value of $F$. Physically, this parameter quantifies the density of spatial settlement at different scales; $F=3$ indicates homogeneous volume growth, $F<3$ reflects the existence of voids and branched structures, and the further $F$ deviates from 3, the more open and patchy the assemblage.

\begin{figure}[!htb]
\centering
\includegraphics[width=1\columnwidth]{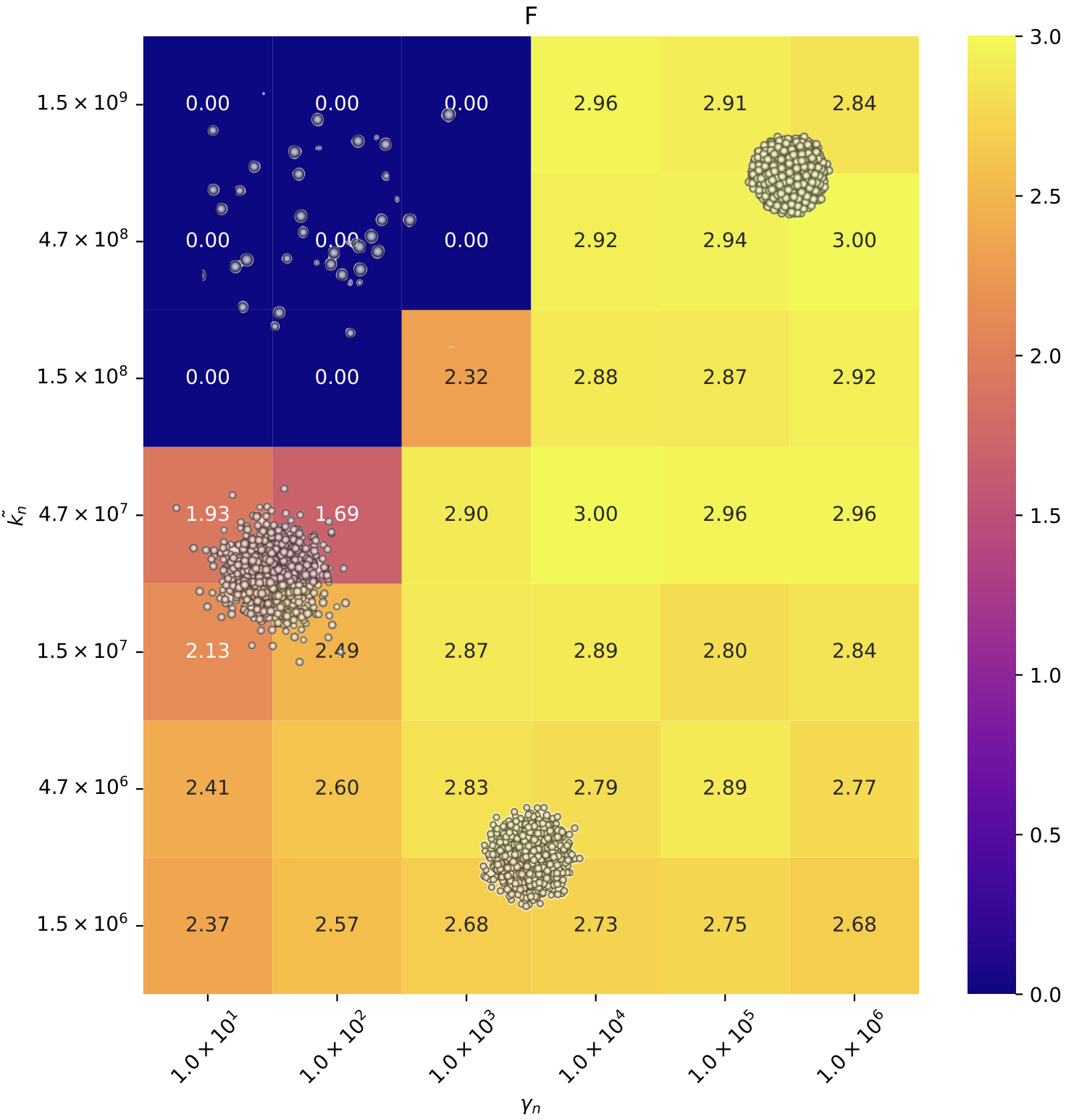}
\caption{Value of fractality $F$ in parametric space ($\widetilde k_n$, $\gamma_n$).}
\label{fig:fractalite_heatmap}
\end{figure}

The heatmap (Fig \ref{fig:fractalite_heatmap}) represents the spatial distribution of $F$ in parameter space ($\widetilde k_n$, $\gamma_n$) in log scale, as well as the final state of some simulations in transparency. The map reveals that for a large value of $\gamma_n$, fractality is close to $3$. $F \approx 3$ implies dense aggregation. If the slope of $\log N(r)$ vs $\log r$ is close to $3$, this means that, by doubling the size of the sphere of radius $r$, the number of particles inside grows as $r^3$. The structure fills three-dimensional space in a volumetric way: we're getting closer to a dense stack or a roughly homogeneous object in 3D. In this context, as the simulation reveals, $F\approx 3$ indicates a highly cohesive aggregate, where particles have few empty spaces between them. This can be associated with low porosity and a high $\phi_f$ volume fraction. For a low value of $\widetilde k_n$, decreasing $\gamma_n$ has the effect of decreasing fractality down to $\sim 2$. $2 \lesssim F < 3$ implies a partially fractal or porous aggregate. When $F$ falls between $2$ and $3$, the growth of $N(r)$ is slower than $r^3$, indicating the presence of voids or internal channels. Typically, a value around $F \approx 2.5$ means that the cluster is relatively dense at the center but thins out towards the periphery, with empty pockets. This intermediate zone corresponds to structures where the aggregate partially clings to the volume, but does not fill it completely.

For a large value of $\widetilde k_n$ and a small $\gamma_n$, the fractality is $0$. This means there has been no aggregation. At low gamma values, increasing $\widetilde k_n$ leads to fractality $F < 2$. Such a value $1 < F \lesssim 2$ implies complex tree-like aggregation. If $F$ goes down to $2$ or less, the aggregate adopts a branched behavior. This is akin to an assembly of branches or filaments. In this case, there are many voids: the energy dissipated at each contact is too low to consolidate the cluster, or the stiffness is too low, allowing particles to adhere at first contact and create long, loosely-packed structures.  

\section{Conclusion}

This campaign of 3D simulations of granular coalescence under gravity has revealed a close link between microscopic contact laws and macroscopic aggregation dynamics. I specify that simulations, focused on grains of the order of a millimeter without van der Waals forces, apply mainly to the terminal phases of gravitational aggregation (decimeter- to meter-sized bodies, "rubble piles"), and are not intended to reproduce the initial stages of sub-cm planetesimal growth, where molecular cohesion dominates.
On the one hand, analysis of the dimensioned aggregation time $\tfrac{T_{\rm agg}}{t_{\rm ff}}$ as a function of the damping coefficient $\gamma_n$ showed a monotonic decrease followed by a plateau, confirming that too little damping keeps particles bouncing, while sufficient damping ($\gamma_n\gtrsim500$) effectively accelerates the formation of a single aggregate without additional gains beyond a certain threshold.
On the other hand, by normalizing contact stiffness by the gravitational scale $k^*=G\,m^2/R^3$, I identified a mechanical optimum around $\widetilde k_n\approx10^6$ : below this point, grains that are too "soft" slip and don't bind (index $C/N_{\rm tot}\approx1$), and above this point, over-elastic rebounds still isolate peripheral particles (rise of $C/N_{\rm tot}$ towards 0.6). Finally, mapping the fractal dimension $F$ in $(\widetilde k_n,\gamma_n)$ space enabled us to link these regimes to microstructure: from $F\approx3$ for compact aggregates to $F\lesssim2$ for branched networks, illustrating the transition from volumetric to fractal.
These results lay the foundations for a unified understanding of gravitational aggregation: dissipation and contact rigidity do not act independently, but combine to determine the rate of formation, the number of final sub-aggregates and the internal structure of the formed object. This quantified connection between microscopic parameters and macroscopic observables opens the way to fine calibration of contact laws for planetesimal modeling and, beyond that, to direct comparison with micro-gravity experiments or N-body simulations coupled with specialized DEMs.

\bibliographystyle{plain}
\bibliography{sample}

\end{document}